\documentclass[aps,prc,twocolumn,superscriptaddress]{revtex4-2}
\usepackage[T1]{fontenc}
\usepackage[utf8]{inputenc}

\usepackage{amsmath,amsfonts,amssymb,graphicx,bm}
\usepackage[colorlinks,linkcolor=blue,citecolor=blue,urlcolor=blue,bookmarks=false]{hyperref}

\usepackage{txfonts}

\usepackage[english]{babel}

\newcommand{\sNN}{\sqrt{s_\text{NN}}}

\newcommand{\vE}{\mathbf{E}}
\newcommand{\vB}{\mathbf{B}}
\newcommand{\vj}{\mathbf{j}}

\renewcommand{\v}{\mathbf{v}}
\newcommand{\x}{\mathbf{x}}
\newcommand{\R}{\mathbf{R}}

\newcommand{\qpi}{\frac{q}{4\pi}}

\begin{document}

\title{Dynamic Calculations of Magnetic Field and Implications on Spin Polarization and Spin Alignment in Heavy Ion Collisions}

\author{Hui Li}
\email{lihui_fd@fudan.edu.cn}
\affiliation{Key Laboratory of Nuclear Physics and Ion-beam Application (MOE), Fudan University, Shanghai, China}

\author{Xiao-Liang Xia}
\email{xiaxl@fudan.edu.cn}
\affiliation{Department of Physics and Center for Field Theory and Particle Physics, Fudan University, Shanghai 200433, China}

\author{Xu-Guang Huang}
\email{huangxuguang@fudan.edu.cn}
\affiliation{Key Laboratory of Nuclear Physics and Ion-beam Application (MOE), Fudan University, Shanghai, China}
\affiliation{Department of Physics and Center for Field Theory and Particle Physics, Fudan University, Shanghai 200433, China}
\affiliation{Shanghai Research Center for Theoretical Nuclear Physics, NSFC and Fudan University, Shanghai 200438, China}

\author{Huan Zhong Huang}
\email{huanzhonghuang@fudan.edu.cn}
\affiliation{Key Laboratory of Nuclear Physics and Ion-beam Application (MOE), Fudan University, Shanghai, China}
\affiliation{Department of Physics and Astronomy, University of California, Los Angeles, CA 90095, USA}

\begin{abstract}
Magnetic field plays a crucial role in various novel phenomena in heavy-ion collisions. We solve the Maxwell equations numerically in a medium with time-dependent electric conductivity by using the Finite-Difference Time-Domain (FDTD) algorithm. We investigate the time evolution of magnetic fields in two scenarios with different electric conductivities at collision energies ranging from $\sNN=$ 7.7 to 200 GeV. Our results suggest that the magnetic field may not persist long enough to induce a significant splitting between the global spin polarizations of $\Lambda$ and $\bar{\Lambda}$ at freeze-out stage. However, our results do not rule out the possibility of the magnetic field influencing the spin (anti-)alignment of vector mesons.
\end{abstract}

\maketitle

\section{\label{sec-intro}Introduction}

In non-central relativistic heavy-ion collisions, two positively charged nuclei collide with non-zero impact parameters, resulting in the generation of a large magnetic field. This magnetic field can reach $10^{18}$ Gauss in Au + Au collisions at $\sNN=200$ GeV at RHIC and $10^{19}$ Gauss in Pb + Pb collisions at $\sNN=5020$ GeV at LHC~\cite{Kharzeev:2007jp,Skokov:2009qp,Voronyuk:2011jd,Bzdak:2011yy,Deng:2012pc,Bloczynski:2012en}. The effect of this strong magnetic field on the Quark-Gluon Plasma (QGP) has attracted much attention due to its potential impacts on many novel phenomena, such as the chiral magnetic effect~\cite{STAR:2021mii,STAR:2022ahj}, the spin polarization of hyperons~\cite{STAR:2017ckg}, the spin alignment of vector mesons~\cite{Acharya:2019vpe,STAR:2022fan}, the charge-dependent directed flow~\cite{STAR:2023jdd}, and the Breit-Wheeler process of dilepton production~\cite{STAR:2019wlg} in heavy-ion collisions.

When making theoretical predictions about the aforementioned effects, a crucial question to be addressed is how the magnetic field evolves over time. In particular, it is important to determine whether the lifetime of the magnetic field is sufficiently long to maintain significant field strength leading to observable effects.

In general, simulations of the magnetic field evolution can be carried out using the following steps. Before the collision, the charge density of the two colliding nuclei can be initialized by utilizing the Wood-Saxon distribution or by sampling the charge position in the nucleus using the Monte-Carlo Glauber model~\cite{Bzdak:2011yy,Bloczynski:2012en}. After the collision, the two charged nuclei pass through each other like two instantaneous currents. Currently, several approaches exist to simulate the collision process. The simplest method is to assume that the two nuclei pass through each other transparently or to incorporate the charge stopping effect using empirical formulae~\cite{Kharzeev:2007jp}. A more sophisticated approach involves simulating the entire collision process through transport models~\cite{Skokov:2009qp,Voronyuk:2011jd,Deng:2012pc}. Once the motion of electric charge is determined, the magnetic field can be calculated using analytical formulae. These methods have been widely employed in previous studies to investigate the evolution of magnetic fields in heavy-ion collisions~\cite{Kharzeev:2007jp,Skokov:2009qp,Voronyuk:2011jd,Bzdak:2011yy,Deng:2012pc,Bloczynski:2012en}.

Previous simulations have shown that the strong magnetic field produced by the colliding nuclei rapidly decays with time in vacuum~\cite{Huang:2015oca}. The lifetime of the magnetic field is primarily determined by fast-moving spectators, and the strong magnetic field only exists during the early stage of the collision. However, the time evolution of the magnetic field can be significantly modified when taking into account the response of the QGP, which is a charge-conducting medium. In this case, when the magnetic field begins to decrease, the induced Faraday currents in the QGP considerably slow down the damping of the magnetic field. Analytical formulae have demonstrated that the damping of the magnetic field in a constant conductive medium can be significantly delayed~\cite{Tuchin:2013apa,Tuchin:2013ie,Gursoy:2014aka,Li:2016tel,Chen:2021nxs}.

However, those analytical formulae only apply to the case of a constant conductivity, which is unrealistic because the conductivity only exists after the collision and the value varies as the QGP medium expands. Therefore, it is essential to numerically calculate the magnetic field. Numerical results can overcome the limitations of analytical calculations and provide unambiguous solutions for time-dependent conductive medium. As a result, numerical results can serve as a more accurate reference for final state observations that are sensitive to the evolution of the magnetic field. It is worth noting that some studies have also simulated the magnetic field by numerically solving the Maxwell equations with an electric conductivity~\cite{McLerran:2013hla,Zakharov:2014dia,Huang:2022qdn} and by combining the magnetic field with the electromagnetic response of the QGP medium~\cite{Roy:2017yvg,Yan:2021zjc,Wang:2021oqq,Grayson:2022asf}.

This paper presents a numerical study of the time evolution of magnetic fields with time-dependent electric conductivities at $\sNN=$ 7.7--200 GeV. To solve the Maxwell equations, we utilize the Finite-Difference Time-Domain (FDTD) algorithm~\cite{Yee:1966FDTD}. The paper is organized as follows: Sec.~\ref{sec-analytical} introduces the analytical formulae. Sec.~\ref{sec-setup} describes the numerical model setup of the charge density, charge current, and the electric conductivity. Sec.~\ref{sec-method} describes the numerical method. Sec.~\ref{sec-result} presents the results and discusses the impact on the spin polarization and the spin alignment. Finally, Sec.~\ref{sec-sum} concludes the results.

\section{\label{sec-analytical}Limitation of analytical formula}

We consider the electromagnetic field which is generated by external current of two moving nuclei and evolves in a conductive medium created in heavy-ion collisions. The electromagnetic field is governed by Maxwell equations:
\begin{align}
    \nabla \cdot \vE  & = \rho,                              \label{Eq1} \\
    \nabla \cdot \vB  & = 0,                                 \label{Eq2} \\
    \nabla \times \vE & = -\partial_t \vB,                   \label{Eq3} \\
    \nabla \times \vB & = \vj + \sigma \vE + \partial_t \vE, \label{Eq4}
\end{align}
where $\rho$ and $\vj$ are the charge density and the charge current, and $\sigma$ is the electric conductivity of the medium.

For a point charge $q$ moving with a constant velocity $\v$, $\rho$ and $\vj$ are:
\begin{align}
    \rho(t,\x) & = q\delta^{(3)}[\x-\x_q(t)],   \\
    \vj(t,\x)  & = q\v\delta^{(3)}[\x-\x_q(t)],
\end{align}
where $\x$ is the position of the field point and $\x_q(t)$ is the position of the point charge at time $t$.

If the conductivity $\sigma$ is a constant, the magnetic field has been rigorously derived as follows~\cite{Tuchin:2013apa,Tuchin:2013ie,Gursoy:2014aka,Li:2016tel,Chen:2021nxs}:
\begin{equation}
    \vB(t,\x) =\qpi\frac{\gamma\v\times\R}{\Delta^{3/2}}\left(1+\frac{\gamma\sigma}{2}|\v|\sqrt{\Delta}\right)e^A,  \label{B_sigma}
\end{equation}
where $\gamma=1/\sqrt{1-\v^2}$ is Lorentz contraction factor, $\R\equiv\x-\x_q(t)$ is the position difference between the field point and the point charge at time $t$, $\Delta\equiv R^2+(\gamma\v\cdot\R)^2$, and $A\equiv -\gamma\sigma(\gamma\v\cdot\R+|\v|\sqrt{\Delta})/2$. If $\sigma$ is set to zero, the above formula can recover to the electromagnetic field in vacuum which can be expressed by the Lienard-Wiechert potential:
\begin{equation}
    \vB(t,\x) =\qpi\frac{\gamma\v\times\R}{[R^2+(\gamma\v\cdot\R)^2]^{3/2}}. \label{L-W}
\end{equation}

Because the Maxwell equations~(\ref{Eq1}--\ref{Eq4}) satisfy the principle of superposition, the formulae~(\ref{B_sigma}) and~(\ref{L-W}) can also be applied to charge distributions rather than just a point charge. Therefore, the formulae have been widely used in the literature~\cite{Kharzeev:2007jp,Skokov:2009qp,Voronyuk:2011jd,Bzdak:2011yy,Deng:2012pc,Bloczynski:2012en,Tuchin:2013apa,Tuchin:2013ie,Gursoy:2014aka,Li:2016tel,Chen:2021nxs} to calculate the magnetic field generated by nuclei in heavy-ion collisions.

However, Eq.~(\ref{B_sigma}) is valid only if 1) the point charge moves with a constant velocity, and 2) the conductivity $\sigma$ is constant (for $t\in [-\infty,\infty]$). Unfortunately, neither of these conditions is realistic in heavy-ion collisions. First, when the collision occurs, charged particles slow down, and the velocities keep changing during the subsequent cascade scattering. Second, the QGP is produced after the collision, which means that the conductivity $\sigma$ is non-zero only after $t = 0$ (the time when the collision happens) and the value of $\sigma$ varies with time.

For these reasons, it is important to develop a numerical method which can solve the Maxwell equations under more complicate and more realistic conditions of $\rho$, $\vj$, and $\sigma$. In this paper, we focus on studying the influence of time-dependent $\sigma$ on the evolution of magnetic field.

\section{\label{sec-setup}Model setup}

\subsection{Charge density and current}

In heavy-ion collisions, the external electric current arises from the contribution of protons in the fast moving nuclei. In this case, we consider two nuclei, which are moving along $+z$ and $-z$ axis with velocity $v_z$, and their projections on the $x$-$y$ plane are centered at $(x=\pm b/2, y=0)$, respectively, with $b$ being the impact parameter.

In the rest frame of a nucleus, the charge distribution can be described by the Wood-Saxon distribution:
\begin{equation}
    f(r) = \frac{N_0}{1+\exp{[(r-R)/a]}}, \label{WS}
\end{equation}
where $R$ is the nuclear radius, $a$ is the surface thickness, and $N_0$ is a normalization factor determined by $4\pi\int f(r)r^2dr = Ze$. Take the gold nucleus as an example, we have $Z=79$, $R=6.38$ fm, $a=0.535$ fm, therefore $N_0\approx 0.0679\ e/\text{fm}^3$.

Then, it is straightforward to derive the charge density and current of the two moving nuclei by a Lorentz boost from Eq.~(\ref{WS}), which leads to
\begin{align}
    \rho^\pm(t,x,y,z) & = \gamma f\left(\sqrt{(x\mp b/2)^2+y^2+\gamma^2(z\mp v_zt)^2}\right), \label{rho}    \\
    j_x^\pm(t,x,y,z)  & = 0, \label{jx}                                                                      \\
    j_y^\pm(t,x,y,z)  & = 0, \label{jy}                                                                      \\
    j_z^\pm(t,x,y,z)  & = \gamma v_z f\left(\sqrt{(x\mp b/2)^2+y^2+\gamma^2(z\mp v_zt)^2}\right), \label{jz}
\end{align}
where the $\pm$ sign over $\rho$ and $j$ on the left side indicates the direction of nucleus' motion along $z$ axis, the velocity $v_z = \sqrt{\gamma^2 - 1} / \gamma$, with $\gamma = \sNN / (2m_\text{N})$ and $m_\text{N}=938$ MeV.

The total charge density and current are given as follows:
\begin{align}
    \rho(t,x,y,z) & = \rho^+(t,x,y,z) + \rho^-(t,x,y,z), \label{total_rho}   \\
    \vj(t,x,y,z)  & = \vj^+(t,x,y,z) + \vj^-(t,x,y,z). \label{total_current}
\end{align}
Eqs.~(\ref{total_rho}) and (\ref{total_current}) can describe the charge and current distributions before the collision exactly when the two nuclei are moving at a constant velocity.

After the collision, the two nuclei are ``wounded'', and some charged particles are stopped to collide with each other. This causes dynamic changes in the charge and current distributions. However, the main goal of this paper is to investigate how the time behavior of the magnetic field is influenced by the time-dependent $\sigma$. As a simplification, we currently assume that the two nuclei pass through each other and continue moving with their original velocity, so the charge and current distributions in Eqs.~(\ref{total_rho}) and (\ref{total_current}) are unchanged after the collision. This allows us to compare our numerical results with the analytical results obtained by Eq.~(\ref{B_sigma}) under the same conditions of $\rho$ and $\vj$, so that we can focus on studying the influence of the time-dependent $\sigma$.

\subsection{Electric conductivity}

Generally, Eq.~(\ref{B_sigma}) is not a realistic description of the electromagnetic response of QGP matter because it assumes a constant conductivity. In reality, the QGP matter exists only after the collision, and the conductivity is time-dependent during the expansion of the system. To provide a more realistic description of the evolution of the magnetic field, it is necessary to consider a time-dependent electric conductivity.

In this study, we consider two scenarios for the electric conductivity. In the first scenario, the conductivity is absent before the collision, and it appears to be constant after the collision. Thus, we can introduce a $\theta({t})$ function to describe it,
\begin{equation}
    \sigma = \sigma_0 \theta(t). \label{case1}
\end{equation}
In this equation, if the constant conductivity $\sigma_0$ were not multiplied by the $\theta(t)$ function, the 
formula~(\ref{B_sigma}) would be valid for calculating the magnetic field. However, as we will show in Sec.~\ref{sec-result}, even with such a minor modification on the electric conductivity, the time behavior of the magnetic field becomes very different.

In the second scenario, we consider the electric conductivity to be absent before the collision, and after the collision the electric conductivity depends on time via
\begin{equation}
    \sigma = \frac{\sigma_0 \theta(t)}{(1 + t / t_0)^{1/3}}. \label{case2}
\end{equation}
The denominator in this equation accounts that the conductivity decreases as the QGP medium expands~\cite{Tuchin:2013ie}. Thus, this scenario provides a more relativistic description of the magnetic field's time behavior in heavy-ion collisions.

\section{\label{sec-method}Numerical method}

In the aforementioned scenarios in Eqs.~(\ref{case1}) and (\ref{case2}), $\sigma$ is time dependent, therefore the analytical results in Eq.~(\ref{B_sigma}) is not applicable, and the Maxwell equations~(\ref{Eq1}--\ref{Eq4}) need to be solved numerically.

Because $\sigma$ is zero before the collision, and the two nuclei move linearly with constant velocity, the electromagnetic field at $t \leq 0$ can be analytically calculated by the Lienard-Wiechert formula as given by Eq.~(\ref{L-W}). This provides the initial condition of the electromagnetic field at $t = 0$.

Once the initial condition is given, the electromagnetic field at $t \geq 0$ is calculated by numerically solving the Maxwell equations~(\ref{Eq1}--\ref{Eq4}). We use the FDTD algorithm~\cite{Yee:1966FDTD} to solve the Maxwell equations. In detail, electric and magnetic fields are discretized on the Yee's grid, and the updating format for $\vB$ and $\vE$ can be constructed by discretizing Eqs.~(\ref{Eq3}) and (\ref{Eq4}) with a finite time step, as follows
\begin{equation}
    \frac{\vB(t + \Delta t) - \vB(t)}{\Delta t} = - \nabla \times \vE\left(t+\frac{\Delta t}{2}\right), \label{B_update}
\end{equation}
and
\begin{eqnarray}
    \frac{\vE(t + \Delta t) - \vE(t)}{\Delta t} & + & \sigma \frac{\vE(t + \Delta t) + \vE(t)}{2} \nonumber \\
    & = & \nabla \times \vB\left(t+\frac{\Delta t}{2}\right) - \vj\left(t+\frac{\Delta t}{2}\right). \label{E_update}
\end{eqnarray}
The Yee's grid provides a high-accuracy method to calculate $\nabla \times \vE$ and $\nabla \times \vB$. As time evolves, $\vE$ and $\vB$ are updated alternately. For example, if $\vB$ is initially known at time $t$ and $\vE$ is initially known at time $t+\Delta t/2$, then one can use the values of $\vE(t+\Delta t/2)$ and Eq.~(\ref{B_update}) to update $\vB$ from $t$ to $t + \Delta t$; and after $\vB(t + \Delta t)$ is obtained, one can use Eq.~(\ref{E_update}) to update $\vE$ from $t + \Delta t/2$ to $t + 3\Delta t/2$. This algorithm provides higher accuracy than the regular first-order difference method.

\section{\label{sec-result}Numerical Results}

Using the numerical method described in Sec.~\ref{sec-method}, we calculate the magnetic field by solving the Maxwell equations~(\ref{Eq1}--\ref{Eq4}) under the conditions of $\sigma = 0$, $\sigma = \sigma_0 \theta(t)$, and $\sigma = \sigma_0 \theta(t)/(1 + t / t_0)^{1/3}$, respectively. As a verification of our numerical method, we have checked that our numerical solution for $\sigma = 0$ matches the analytical result by Eq.~(\ref{L-W}). We also calculate the magnetic field under the condition of $\sigma = \sigma_0$ using the analytical formula~(\ref{B_sigma}) for comparison. In all the results presented in this section, the values of $\sigma_0$ and $t_0$ are set to be $\sigma_0= 5.8$ MeV and $t_{0} = 0.5$ fm/c, which are taken from Ref.~\cite{Tuchin:2013ie}.

\subsection{$\sigma = \sigma_0 \theta(t)$ vs $\sigma = \sigma_0$}

\begin{figure*}[t]
    \includegraphics[scale=0.8]{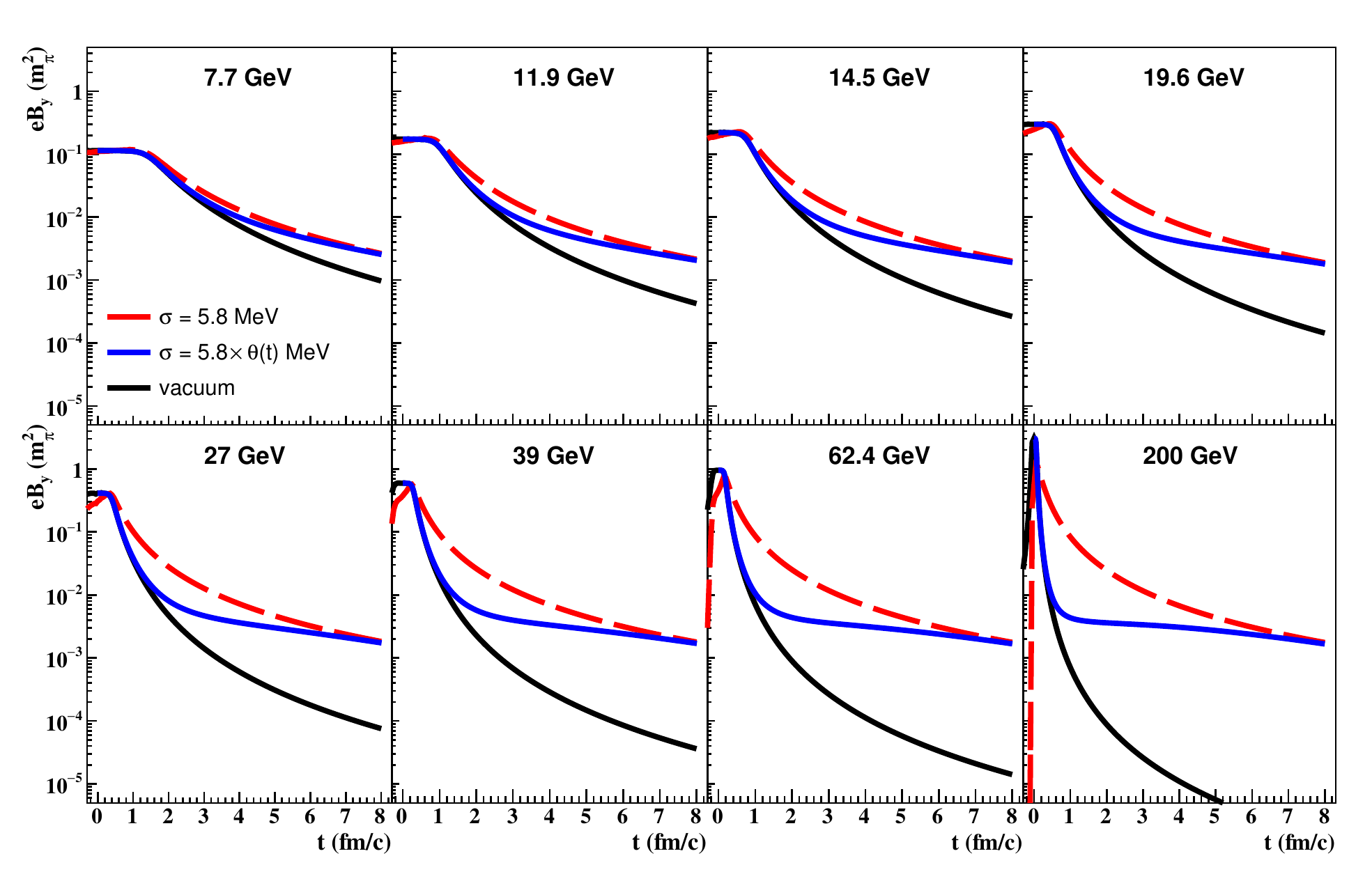}
    \caption{The time evolution of magnetic field $B_y$ in Au + Au collisions at $\sNN=$ 7.7--200 GeV with impact parameter $b = 7$ fm. The results of $B_y$ from the numerical algorithm [blue curve, for $\sigma = \sigma_0\theta(t)$] and from the analytical formula given by Eq.~(\ref{B_sigma}) (red dashed curve, for $\sigma = \sigma_0$) are shown. The evolution of $B_y$ in vacuum ($\sigma = 0$, black curve) is also shown as a baseline.}
    \label{By_compare1}
\end{figure*}

Figure~\ref{By_compare1} displays the time evolution of the magnetic field in the out-of-plane direction ($B_y$) at the center of collision ($\mathbf{x} = 0$) in Au+Au collisions for energies ranging from 7.7 to 200 GeV with impact parameter $b = 7$ fm. The results of $\sigma = \sigma_0 \theta(t)$ are calculated using the numerical algorithm described in Sec.~\ref{sec-method}, while the results of $\sigma = \sigma_0$ are calculated using the analytical formula given by Eq.~(\ref{B_sigma}). The magnetic field in vacuum ($\sigma = 0$) is also shown as a baseline.

In general, the presence of electric conductivity delays the decreasing of the magnetic field. However, the time behavior of the magnetic field under the condition of $\sigma = \sigma_0 \theta(t)$ is very different from that of $\sigma = \sigma_0$.

In Figure~\ref{By_compare1} we can see that, in the case of $\sigma = \sigma_0$ (namely, $\sigma$ is constant at both $t < 0$ and $t > 0$), the magnitude of magnetic field is different from the vacuum baseline since a very early time. On the other hand, in the case of $\sigma = \sigma_0 \theta(t)$, the difference between the magnetic field and the vacuum baseline is negligible at early time stages ($t < 1$ fm/c for 200 GeV or $t < 3$ fm/c for 7.7 GeV). This is because that $\sigma$ exists only after the collision and it needs some time to build the effect on delaying the magnetic field's decay. Only at very late time stage ($t > 7$ fm/c), when the evolution system has ``forgotten'' whether $\sigma$ is zero or not before $t = 0$, the curves of $\sigma = \sigma_0 \theta(t)$ and of $\sigma = \sigma_0$ converge. In the middle time stage, the magnitude of the magnetic field is ranked in the order: $B[\text{vacuum}] < B[\sigma=\sigma_0\theta(t)] < B[\sigma=\sigma_0]$.

Our results indicate that the analytical formula~(\ref{B_sigma}) significantly overestimates the magnetic field in the early and middle time stage compared to the numerical results. The difference between the analytical and numerical results arises from the $\theta(t)$ function introduced in Eq.~(\ref{case1}). It is important to note that the conductivity is absent at $t < 0$ in realistic collisions, therefore the formula~(\ref{B_sigma}) is not applicable. This remarks the importance of considering time-dependent $\sigma$ and solving the Maxwell equations numerically. At the late time stage, although the analytical results agree well with the numerical ones, the magnetic field has become very small and has little impact on final observables.

\subsection{$\sigma = \sigma_0 \theta(t)$ vs $\sigma = \sigma_0 \theta(t) / (1 + t / t_0)^{1/3}$}

\begin{figure*}[t]
    \includegraphics[scale=0.8]{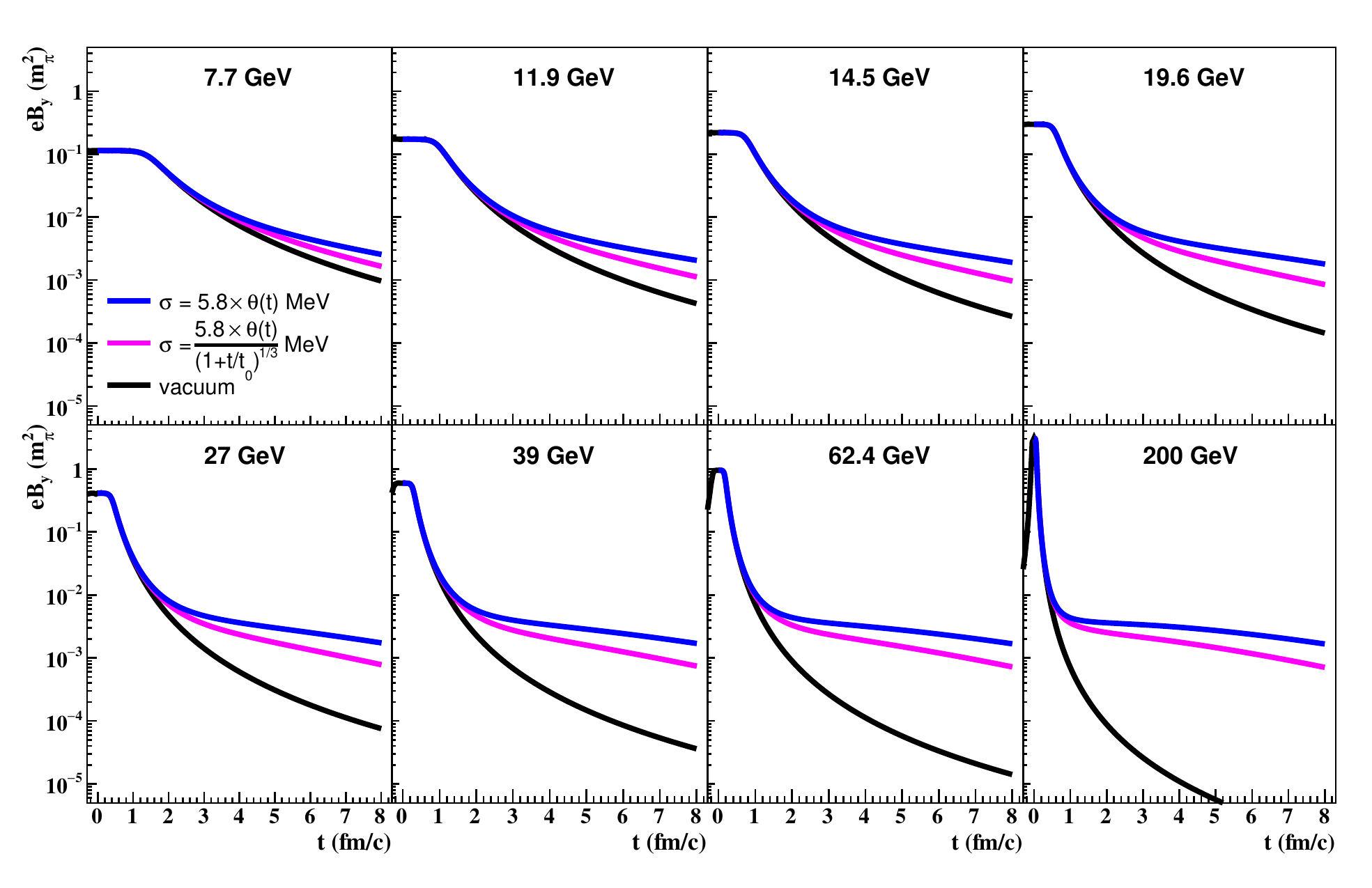}
    \caption{The time evolution of magnetic field $B_y$ in Au + Au collisions at $\sNN=$ 7.7--200 GeV with impact parameter $b = 7$ fm. The results are calculated numerically under the conditions of $\sigma = \sigma_0 \theta(t)$ (blue curve), $\sigma = \sigma_0 \theta(t) / (1 + t / t_0)^{1/3}$ (magenta curve), and $\sigma=0$ (black curve), respectively.}
    \label{By_compare2}
\end{figure*}

The electric conductivity in heavy-ion collisions is a time-dependent quantity due to the expansion of the QGP. Therefore, we consider a more realistic scenario where the electric conductivity decreases with time as given by Eq.~(\ref{case2}). Figure~\ref{By_compare2} shows the corresponding results, which are compared to the results under the conditions of $\sigma = \sigma_0 \theta(t)$ and $\sigma = 0$.

We see again that, in both scenarios of $\sigma = \sigma_0 \theta(t)$ and of $\sigma=\sigma_0\theta(t) / (1 + t / t_0)^{1/3}$, the magnitude of the magnetic field  does not obviously diverge from the vacuum baseline at the early time stage. At later time, the differences is manifested, and we see that $B[\text{vacuum}] < B[\sigma=\sigma_0\theta(t) / (1 + t / t_0)^{1/3}] < B[\sigma=\sigma_0\theta(t)]$.

Needless to say, the decreasing conductivity has smaller effect on delaying the magnetic field's decay than a constant one. Nevertheless, Figure~\ref{By_compare2} shows that the magnitude of the magnetic field with $\sigma=\sigma_0\theta(t) / (1 + t / t_0)^{1/3}$ are more close to the one of $\sigma=\sigma_0\theta(t)$ than to the vacuum baseline, especially at high energies. This suggests that the even if the conductivity decreases, it still has an obvious effect on delaying the damping of the magnetic field. However, this effect is only significant in late time stage, when the magnetic field has already decreased.

\subsection{Impact on the spin polarization}

Now let us discuss the impact of the magnetic field on the splitting between the global spin polarizations of $\Lambda$ and $\bar{\Lambda}$. The magnetic-field-induced global spin polarization of $\Lambda$ and $\bar{\Lambda}$ can be calculated using the following formula~\cite{Becattini:2016gvu}
\begin{equation}
    P_{\Lambda/\bar{\Lambda}} = \pm\frac{\mu_{\Lambda}B}{T}, \label{P_Lambda}
\end{equation}
where $\mu_{\Lambda}$ is the magnetic moment of $\Lambda$ and is equal to $-0.613\mu_{N}$, with $\mu_{N}$ being the nuclear magneton, and $T$ is the temperature when the hyperon spin is ``freezed''. We shall use the hadronization temperature $T\approx 155$ MeV as an estimate. Then the splitting between the $\Lambda$ and $\bar{\Lambda}$ global spin polarizations is given by
\begin{equation}
    P_{\bar{\Lambda}}-P_{\Lambda} = 0.0826\frac{eB}{m_{\pi}^2}.
    \label{difference}
\end{equation}

Based on the numerical results presented in Figure~\ref{By_compare2}, the magnitude of the magnetic field at late time is of the order of $eB_y\sim 10^{-3}$ -- $10^{-2}\ m_{\pi}^2$, which is significantly smaller than the initial values at $t=0$. Therefore, the effect of the magnetic field on the global spin polarizations of $\Lambda$ and $\bar{\Lambda}$ is negligible, as the splitting can be no larger than $0.1\%$. This is consistent with the recent STAR data~\cite{STAR:2023ntw} which puts an upper limit of $P_{\bar{\Lambda}}-P_{\Lambda} < 0.24\%$ at $\sNN=$ 19.6 GeV and $P_{\bar{\Lambda}}-P_{\Lambda} < 0.35\%$ at $\sNN=$ 27 GeV. In conclusion, our results suggest that the magnetic field is not sufficiently long-lived to provide a distinguishable splitting between the $\Lambda$ and $\bar{\Lambda}$ global spin polarizations under the current experimental accuracy; similar results were obtained also in Ref.~\cite{Peng:2022cya}.

\subsection{Impact on the spin alignment}

The magnetic field also plays an important role in the spin (anti-)alignment of vector mesons. For vector mesons such as $\phi$ and $K^{*0}$, the spins of the constituent quarks in the meson have a lager chance to be anti-algined [i.e.~the $(|\uparrow\downarrow\rangle+\vert\downarrow\uparrow\rangle)/\sqrt{2}$ state] than to be aligned ($|\uparrow\uparrow\rangle$ or $|\downarrow\downarrow\rangle$ state) in an external magnetic field~\cite{Liu:2014ixa,Yang:2017sdk}. This effect can be explored experimentally by measuring the spin-density matrix element $\rho_{00}$. We note that $\rho_{00}$ is a frame dependent quantity. The following formulae show the $\rho_{00}$ with respect to $x$, $y$, and $z$ axis, respectively~\cite{Efremov:1981vs,Xia:2020tyd}:
\begin{align}
    \rho_{00}^{(x)} & = \frac{1-P_x^qP_x^{\bar{q}}+P_y^qP_y^{\bar{q}}+P_z^qP_z^{\bar{q}}}{3+\mathbf{P}_{q}\cdot\mathbf{P}_{\bar{q}}}, \label{spin-alignment1} \\
    \rho_{00}^{(y)} & = \frac{1-P_y^qP_y^{\bar{q}}+P_x^qP_x^{\bar{q}}+P_z^qP_z^{\bar{q}}}{3+\mathbf{P}_{q}\cdot\mathbf{P}_{\bar{q}}}, \label{spin-alignment2} \\
    \rho_{00}^{(z)} & = \frac{1-P_z^qP_z^{\bar{q}}+P_x^qP_x^{\bar{q}}+P_y^qP_y^{\bar{q}}}{3+\mathbf{P}_{q}\cdot\mathbf{P}_{\bar{q}}}. \label{spin-alignment3}
\end{align}
where $(P_x^q,\ P_y^q,\ P_z^q)$ and $(P_x^{\bar{q}},\ P_y^{\bar{q}},\ P_z^{\bar{q}})$ are spin polarization vectors of the constituent quark and anti-quark, respectively.

Our results have shown that the global spin polarization induced by the magnetic field is a small amount ($<0.1\%$), therefore one may expect that the contribution from the magnetic field to the spin alignment (measured via $\rho_{00}-1/3$, which is proportional to the square of the magnetic field) will be even smaller. However, it should be realized that our calculations do not take into account the fluctuations in the charge density and current. Therefore, the results should be interpreted as the averaged magnetic field, which suggest that the average values such as $\langle P_{q} \rangle$ and $\langle P_{\bar{q}} \rangle$ are small, but do not imply that the correlation between $P_{q}$ and $P_{\bar{q}}$ is small. Instead, when a vector meson is formed by combination of a quark and an anti-quark, the distance between the quarks should be small enough, thus $P_{q}$ and $P_{\bar{q}}$, which arise from the fluctuation of magnetic field, are highly correlated. This can lead to a massive contribution to $\rho_{00}$.

Therefore, our results do not rule out the possible effect of the magnetic field on the spin (anti-)alignment of vector mesons. For the same reason, the spin alignment of vector mesons can also arise from the fluctuation of other fields such as vorticity~\cite{Xia:2020tyd}, temperture gradient~\cite{Becattini:2016gvu}, shear tensor~\cite{Becattini:2021suc,Liu:2021uhn}, and strong-force field~\cite{Sheng:2019kmk,Sheng:2020ghv,Kumar:2023ghs}. Finally, it is important to note that, if the spin alignment is mainly contributed by fluctuations, then the value of $\rho_{00}$ is not constrained by the value of global or local $\Lambda$ polarizations. This may explain the significant value of $|\rho_{00}-1/3|$ in the experimental data~\cite{Acharya:2019vpe,STAR:2022fan}, whereas the global or local $\Lambda$ polarizations are much smaller~\cite{STAR:2017ckg,Adam:2019srw,Acharya:2019ryw,ALICE:2021pzu,STAR:2023ntw}.

\section{\label{sec-sum}Summary}

In this study, we present a numerical method to solve the Maxwell equations and investigate the evolution of magnetic field in heavy-ion collisions. We also discuss the impact of the magnetic field on the spin polarizations of $\Lambda$ and $\bar{\Lambda}$ as well as the spin alignment of vector mesons.

We demonstrate that although the electric conductivity can delay the decay of the magnetic field, this effect has been overestimated by the analytical formula which assumes a constant conductivity. After taking into account that the conductivity only exists after the collision, we find that the magnetic field is not sufficiently long-lived to induce a significant splitting between the global spin polarizations of $\Lambda$ and $\bar{\Lambda}$.

On the other hand, the spin alignment of vector meson is a measure of correlation between the spin polarizations of quark and anti-quark, instead of the spin polarization being squared solely. Therefore, although the averaged spin polarization induced by the magnetic field is very small, our results do not rule out the possibility that the fluctuations of the magnetic field, as well as other fields, can have a significant contribution to the spin alignment of vector meson.

\begin{acknowledgments}

    We thank Dmitri Kharzeev and Oleg Teryaev for useful comments on the retreat on Spin Dynamics, Vorticity, Chirality and magnetic field workshop. This work was supported by the NSFC through Grants No.~11835002, No. 12147101, No. 12225502 and No. 12075061, the National Key Research and Development Program of China through Grant No. 2022YFA1604900, and the Natural Science Foundation of Shanghai through Grant No. 20ZR1404100. H. L was also supported by the China Postdoctoral Science Foundation 2019M661333.

\end{acknowledgments}

\bibliographystyle{apsrev4-2}
\bibliography{ref.bib}

\end{document}